\documentclass[osajnl,twocolumn,showpacs,superscriptaddress,10pt]{revtex4-1}

\usepackage{amsmath,amssymb,graphicx}

\begin{document}

\title{All-Optical Detection of Acoustic Pressure Waves with applications in Photo-Acoustic Spectroscopy}

\author{Philip G. Westergaard}
\affiliation{Danish Fundamental Metrology, Matematiktorvet 307, DK-2800 Kgs. Lyngby, Denmark}

\author{Mikael Lassen}\email{Corresponding author: ml@dfm.dk}
\affiliation{Danish Fundamental Metrology, Matematiktorvet 307, DK-2800 Kgs. Lyngby, Denmark}

\begin{abstract}
An all-optical detection method for the detection of acoustic pressure waves is demonstrated. The detection system is based on a stripped (bare) single-mode fiber. The fiber vibrates as a standard cantilever and the optical output from the fiber is imaged to a displacement-sensitive optical detector. The absence of a conventional microphone makes the demonstrated system less susceptible to the effects that a hazardous environment might have on the sensor. The sensor is also useful for measurements in high temperature (above $200^{\circ}$C) environments where conventional microphones will not operate. The proof-of-concept of the all-optical detection method is demonstrated by detecting sound waves generated by the photo-acoustic effect of NO$_2$ excited by a 455 nm LED, where a detection sensitivity of approximately 50 ppm was achieved.
\end{abstract}

\maketitle

\section{Introduction}

Trace gas sensors for hazardous and hostile environments are of increasing importance for industrial monitoring \cite{Hodgkinson2013,Sigrist2008}. To this end, photo-acoustic (PA) sensors are good candidates as they are both versatile and sensitive instruments \cite{Harren2000,Tam1986,lassen2014,Saarela2011,Lassen2015,Peltola2015}. The PA technique is based on the detection of acoustic pressure waves (sound) that are generated due to molecular absorption of modulated optical radiation \cite{PAS1881}. A typical experimental setup for PA involves a light source that is either mechanically chopped or modulated in current, and an absorption cell with a microphone or any other kind of pressure sensitive detector. Conventionally, a standard electret microphone is used to monitor the acoustic waves that appear after the radiation is absorbed. However, a PA sensor without the need for an electrical microphone would be particularly useful where the measurement environment is hostile. Detection via optical methods eliminates the need for detection devices that may create hazardous situations, such as placing electrical wiring in a flammable environment, or where contamination from foreign particles or contaminants, such as solder, may be a problem. Compared to the conventional electronic microphone, fiber-optic acoustic sensors also have the advantages of large bandwidth, as well as immunity to electromagnetic and radio frequency interference.  A number of detection schemes that take advantage of an all-optical sound detection have already been investigated \cite{Cole1977,Koskinen2008,Guo2014,Jin2015,Torras-Rosell2012,Cao2013}. In comparison with other all-optical sound detection techniques our fiber sensor is in addition to being useful in hazardous and hostile environments also relatively simple in optical design and alignment and cheap to build, which makes the PA sensor very attractive for commercial applications.

We demonstrate an all-optical pressure sensor where the detection system is based on a stripped (bare) single-mode fiber, which eliminates the environmental limitation of a standard microphone. The fiber is located inside the PA cell and vibrates as a cantilever when excited by acoustic pressure waves. The vibration is enhanced when the excitation frequency is close to the eigenfrequency of the fiber. A seed laser is transmitted through the fiber and used to monitor the vibration by imaging the output light from the fiber tip onto a displacement/tilt sensitive detector, thus producing a signal proportional to the concentration of the molecules under investigation by virtue of the PA effect. An explicit PA measurement on NO$_2$ excited by a 455 nm LED is demonstrated.

\section{Experimental setup}
\label{sec:examples}
\begin{figure}[h!]
\centerline{\includegraphics[width=.95\columnwidth]{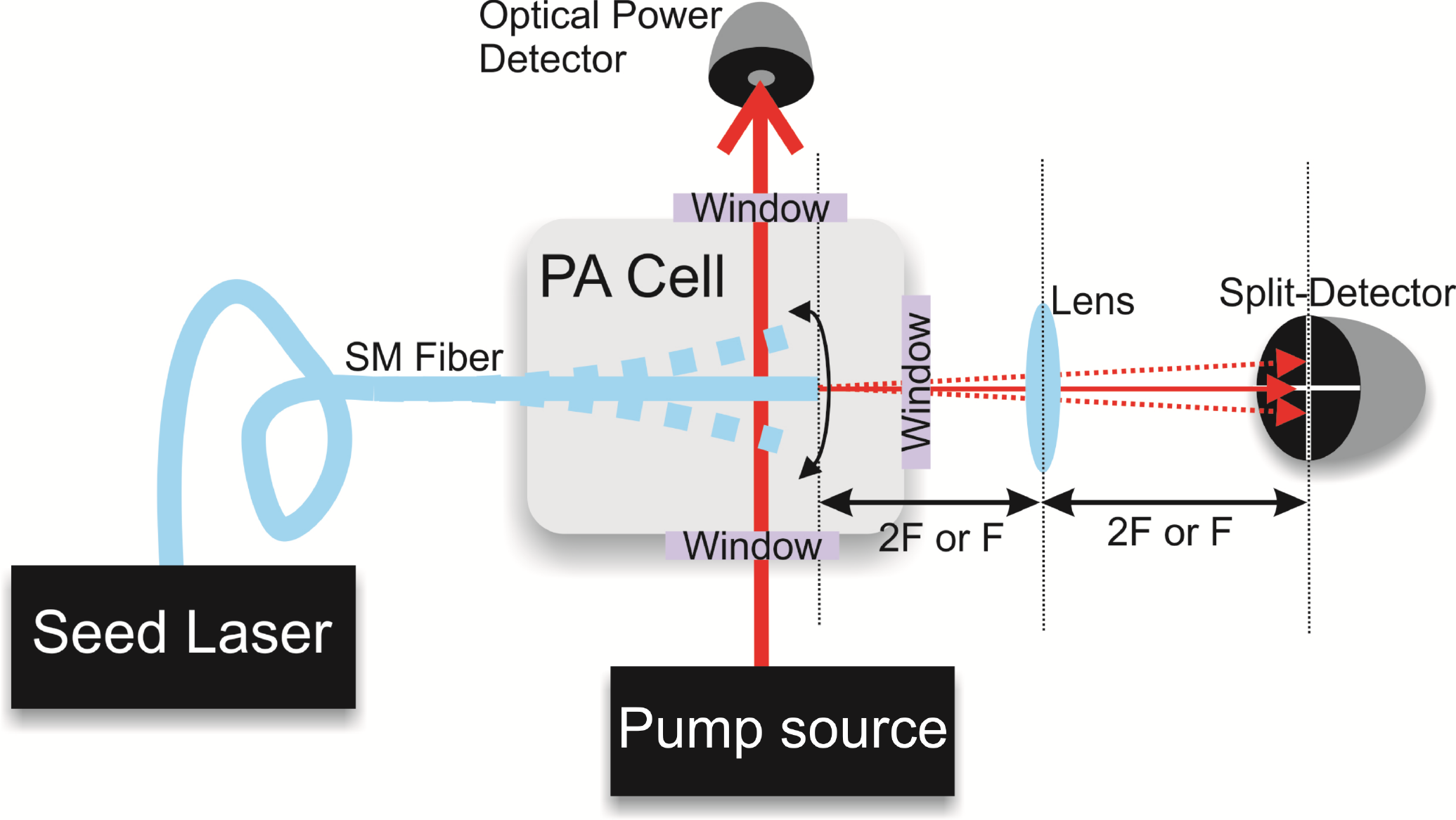}}
\caption{Schematics of the all-optical detection method. The split-detector used here is a Quadrant Position Detector (PDQ30C from Thorlabs) for measuring displacements in both the horizontal and vertical direction. A lens is used to image the tip of the fiber onto the split-detector in the near-field (NF) or far-field (FF) or any super-position which maximizes the signal. The windows have high transmission for the seed and pump light. The optical power detector is used to monitor the optical pump power. The windows ensure that the pump light can enter the PA cell, interact with the molecules and then exit the PA cell. The third window is used for the seed laser.}
\label{fig1}
\end{figure}

Figure \ref{fig1} depicts the schematics of the fiber-based all-optical acoustic detection system.
A standard single-mode optical fiber is used as the sound transducer. The optical fiber is stripped in order to remove the protective polymer coating surrounding the optical fiber. The stripped fiber is placed in a vacuum tight cell with two optical windows for the pump laser and one optical window for the output of the seed laser, with the seed fiber inserted opposite this window in the cell. We use a fiber-coupled DFB laser giving 1.5 mW at 1310 nm (Thorlabs S3FC1310) as seed laser. An LED or other light sources could also be used for the seed light, provided enough light can be coupled into the seed fiber. Depending on the detector, reducing the optical power could reduce the obtainable SNR. The pump light source is tuned to a specific molecular absorption line corresponding to the particular molecular gas under investigation. Absorption causes local heating and results in thermal expansion (pressure waves) of the molecules. The resulting vibration of the fiber is then monitored via the seed laser from the fiber. The movement of the fiber generates a displacement and tilt of the seed laser beam \cite{Delaubert2006}. The conventional way to measure the displacement of a laser beam is to use a split detector as shown in Fig.~\ref{fig1}. This displacement/tilt technique has traditionally be used for many different applications including: optical tweezers, atomic force microscopes, beam positioning for gravitational wave detectors and satellite alignment. The split-detector used here is an InGaAs quadrant position detector (PDQ30C from Thorlabs). By changing the imaging properties from near field (NF) to far field (FF), the displacement and tilt, respectively, of the fiber tip is imaged on the detector \cite{Delaubert2006}. This Fourier transform relation is a well-known result in classical optics, for which a displacement in the focal plane of a simple lens is changed into an inclination relative to the propagation axis in the far field. The difference between the intensity on each section of the split detector yields a photocurrent proportional to the displacement, and therefore proportional to the acoustic sound pressure. Experimentally, the fiber tip will most likely not oscillate exclusively horizontally or vertically, and we therefore just choose the direction on the quadrant detector with the largest signal. After this, the imaging properties are changed (i.e., the lens is moved) to further maximize the signal. Generally, the position of the lens that maximizes the oscillation signal will not correspond purely to a displacement or tilt, but rather a mix of the two.

\section{Results}
\subsection{Comparison with a standard microphone}
\begin{figure}[h!]
\centerline{\includegraphics[width=.95\columnwidth]{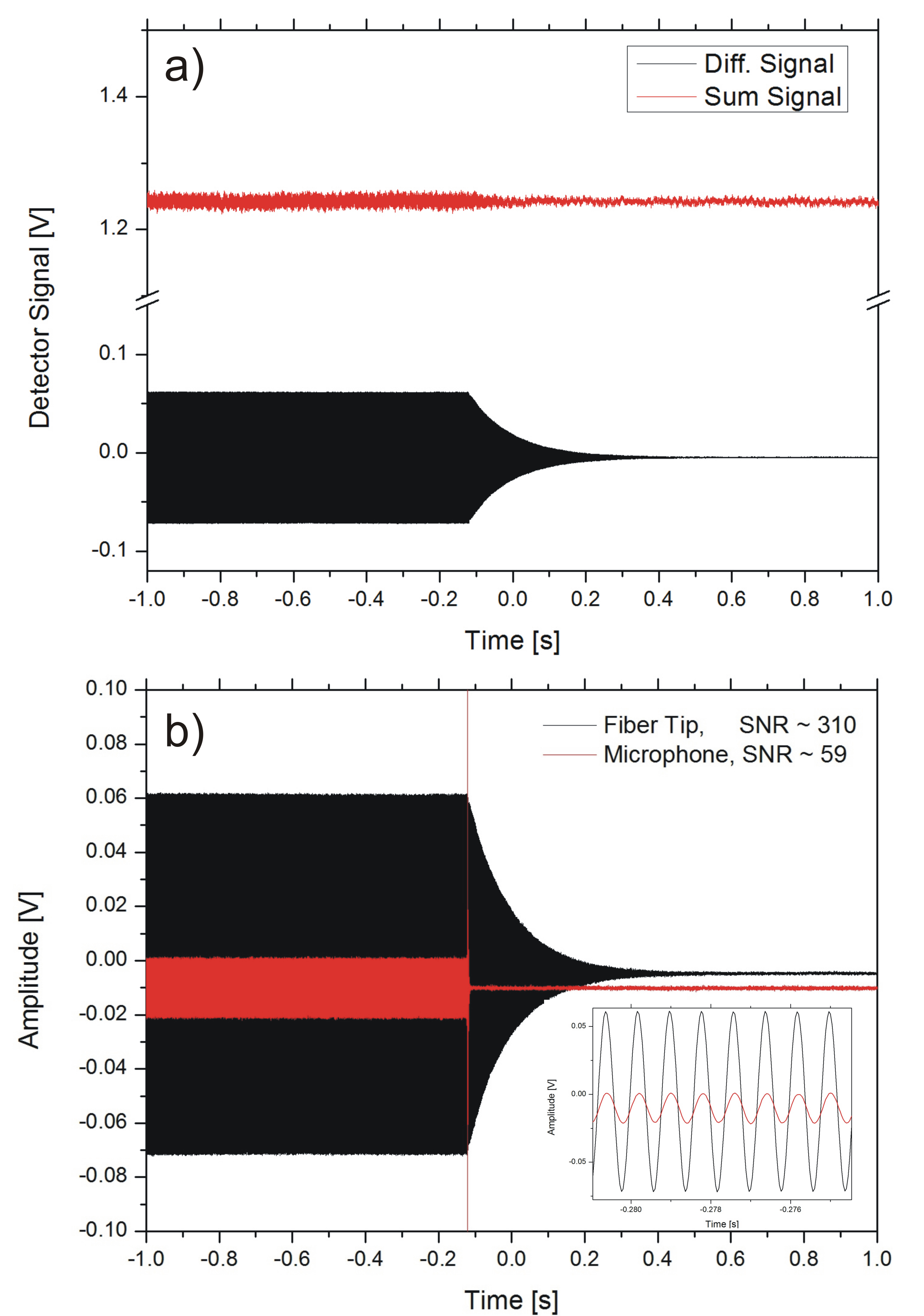}}
\caption{Comparison of SNR for the all-optical sensor and a standard electret microphone. The experimental data was made with a loudspeaker at around 1 kHz as sound source, which was switched off at around t = -0.1 s. The fiber used as the transducer was a standard SM-fiber approximately 10 mm long. a) Sum and difference photo-currents from the split-detector. b) Difference signal (same as in figure a)) for the all optical sensor (black curve) and the signal from a standard electret condenser microphone (red curve). The inset shows a zoom of the traces, where it is seen that the monitored signals are given by pure harmonic oscillations.}
\label{fig2}
\end{figure}

In order to test the fiber cantilever sensor and characterize it, we use a loudspeaker to induce acoustic pressure waves and an electret condenser microphone for benchmarking as depicted in Fig. \ref{fig2}. In Fig. \ref{fig2} the sum and difference signal from the detector is shown.
We find that besides measuring the displacement/tilt of the seed laser the difference signal also subtracts classical noise from the seed laser, thus enhancing the signal-to-noise ratio (SNR) of the optical measurement, which for high common mode rejection can therefore be conducted at the shot-noise limit and even below the shot-noise limit \cite{Lassen2007}. For the data shown Figure \ref{fig2}a) the limit is set by electronical noise of the detector. In Figure \ref{fig2}b) the signal from the difference between the intensity in each section of the split detector is shown when excited by a loudspeaker. The black curve shows the response of the fiber sound transducer and the red curve shows the response from electret condenser microphone (KEEG1538WB-100LB, sensitivity: -42 dB and 58 dB SNR). The acoustic sound wave was generated with a frequency of approximately 1 kHz corresponding to the eigenfrequency for a 10 mm SM-fiber. It can be seen that the fiber cantilever sensor has a response SNR approximately 5 times larger than the microphone at 1 kHz. An advantage of the fiber sensor is that the resonance frequency can simply be tuned by changing the length of the fiber, thus rendering miniaturization and lab-on-a-chip solutions easier to obtain.

\subsection{Vibration frequency}
The fiber vibration can be seen as a cantilever beam system where one end of the fiber is rigidly fixed to a support and the other end is free to move. The natural (eigen) frequency of a 1-dimensional vibration is given by \cite{Han1999},
\begin{equation}
F[L,n] = \frac{k*\alpha^2_n}{L^2},
\label{eq1}
\end{equation}
where the constant $k$ depends on the elastic modulus, the mass of the fiber (cantilever) and the area moment of inertia. $L$ is the length of the fiber and $\alpha_n$ is a constant for the oscillation mode $n$, which for the first modes are given by 1.875, 4.694 and 7.855, respectively.
\begin{figure}[h!]
\centerline{\includegraphics[width=.95\columnwidth]{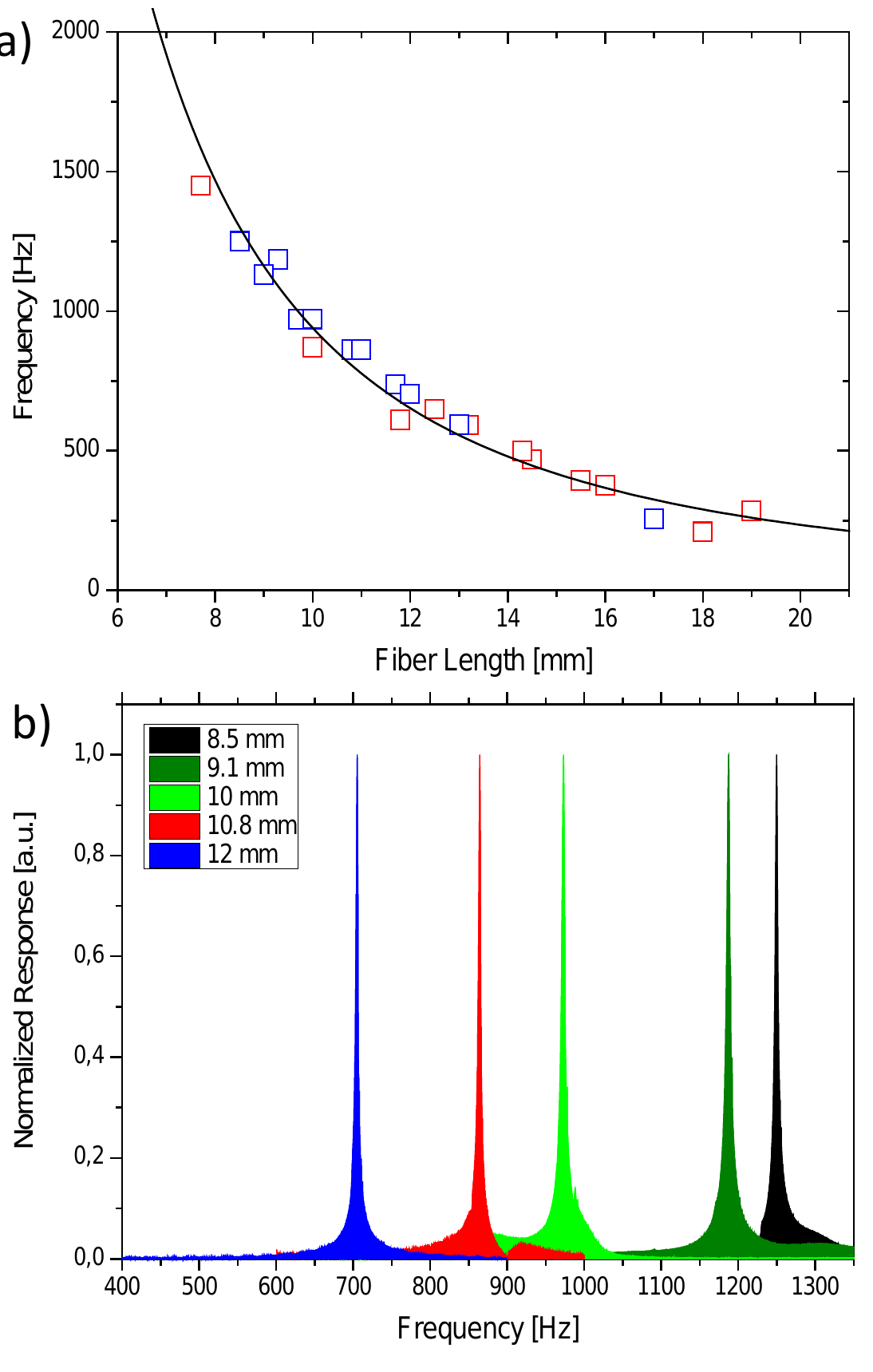}}
\caption{a) Eigenfrequency of the fiber vibration as a function of fiber length. The squares are the experimental data - the red and blue squares are measured with the SM-fiber and the HC-PCF, respectively. The black curve is a fit to the data using the theoretical model in \eqref{eq1}. The uncertainty on the fiber length measurements is approximately 5$\%$. b) Normalized amplitude response as function of frequency for five different fiber lengths. The fiber used for these measurements was the SM fiber together with a compact DFB laser at 1310 nm. The quality factor of the fiber vibration is approximately 220 and is roughly the same for different fiber lengths in the range from 8 mm to 14 mm. The experimental data was made with a loudspeaker as acoustic excitation source.}
\label{fig3}
\end{figure}	
In Fig. \ref{fig3}a) the experimentally measured natural frequency as function of length is plotted for various fiber lengths and two different fibers: a single mode fiber (SM fiber) with high transmission for 980-1550 nm (P1-980A-FC-1 from Thorlabs) and a hollow-core photonic bandgap fiber (HC-PCF-1060 from NKT Photonics). The diameter of the two stripped fibers is the same ($125\; \mu$m). The two fibers have the same frequency response as a function of length despite being structurally different. We therefore choose the cheaper SM fiber as the acoustic sound transducer. The black curve in Fig. \ref{fig3}a) is the simple natural frequency of a 1-dimensional vibration and it can be seen that good agreement is found between the simple theory and the experimentally measured frequencies within the uncertainties of the length measurements. The natural frequency also scales with the diameter $d$ of the fiber as $d^2$ \cite{Han1999}. For simplicity, we keep the diameter constant and vary only the fiber length. In Fig. \ref{fig3}b) the normalized response as a function of frequency for five different fiber lengths are shown. We find that for a SM-fiber with lengths between 8.5 mm and 12 mm the quality-factor is approximately 220 and does not change with length over this region. For the PA experiment a good choice for fiber length would be around 8-10 mm. This also means that the PA cell size in principle can have a minimum of 10x3x3 mm$^3$. Small volumes are not only important for lab-on-a-chip-based solution, but also an advantage for the PA signal since the generated acoustic pressure amplitude is inversely proportional to the volume of the cell \cite{Harren2000}.

\subsection{Temperature Tests}
\begin{figure}[h!]
\centerline{\includegraphics[width=.95\columnwidth]{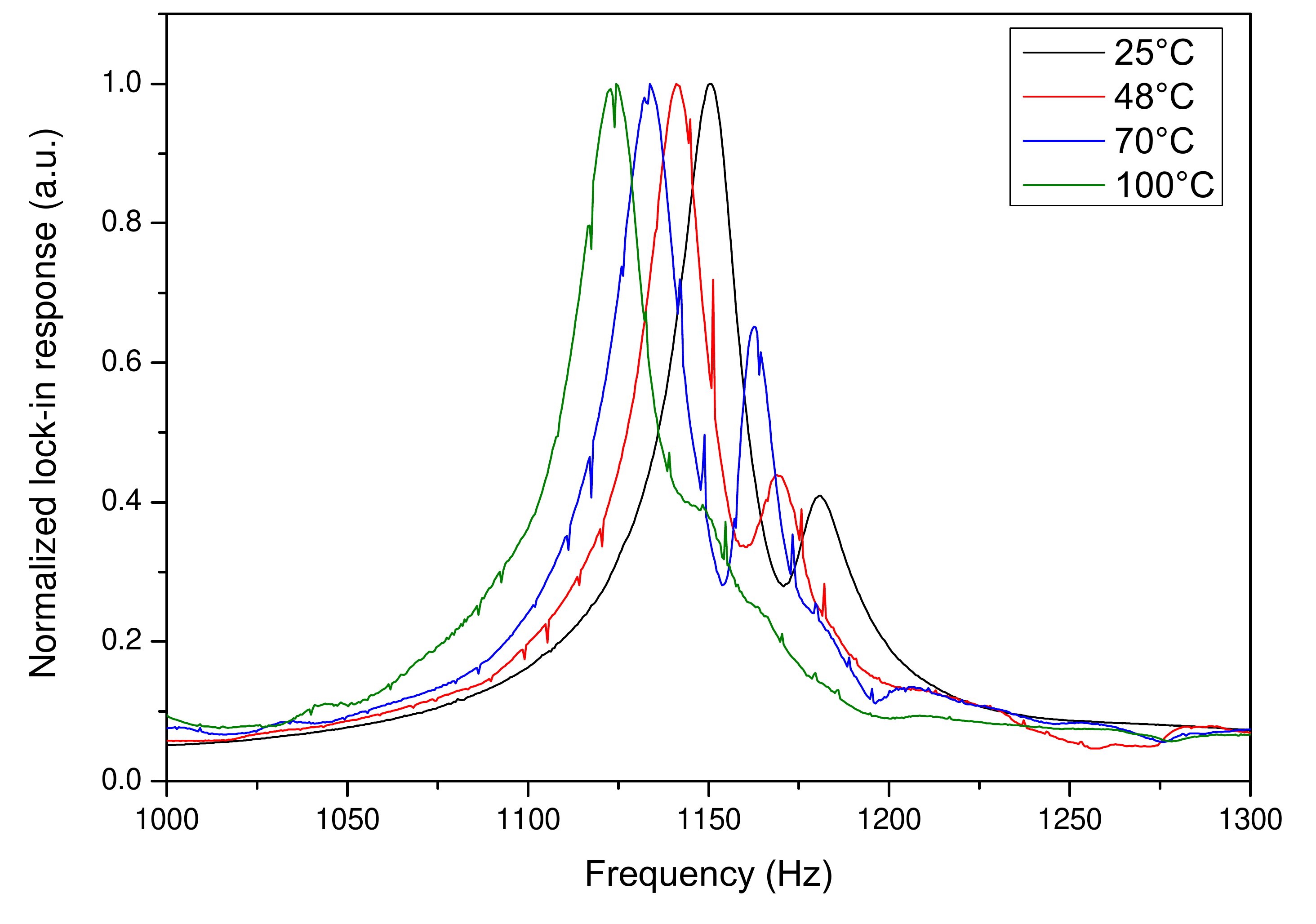}}
\caption{The response of the test system when the cell temperature is changed. Both the shape of the mechanical resonance of the fiber cantilever and the Q-factor is maintained. By changing the temperature from 25 to 100 $^\circ$C a shift in eigenfrequency is observed. The observed decrease in frequency with increase in temperature is around 0.55 Hz/$^\circ$C.  }
\label{fig4}
\end{figure}

To support the claim that the system can endure higher temperatures than microphone-based systems, tests have been performed up to 100 $^\circ$C. The limiting factor in terms of temperature was the heating source (a resistive heating element mounted on the outside of the PA cell), whereas the fiber sensor most likely can be operated up to more than 200-300$^\circ$C.  The normalized lock-in response of the system for four different cell temperatures is shown in Figure \ref{fig4}. The first thing to note is the SNR which remains the same for the different temperatures. Also, the resonance frequency of the fiber tip is slightly decreased as the temperature is increased. This is likely due to slight changes in the way the fiber is fixed. For this test the fiber was glued with fast curing epoxy inside the PA cell. Note that for standard PA sensor an increase in temperature from 25$^\circ$C to 75$^\circ$C would increase the resonance frequency of the PA sensor with 12$\%$. This has the advantage compared to standard PA sensors that our fiber sensor in principle does not need any re-calibration of the light source modulation frequency when the temperature is changed or fluctuating.

\subsection{PA detection of NO$_2$}
\begin{figure}[h!]
\centerline{\includegraphics[width=.95\columnwidth]{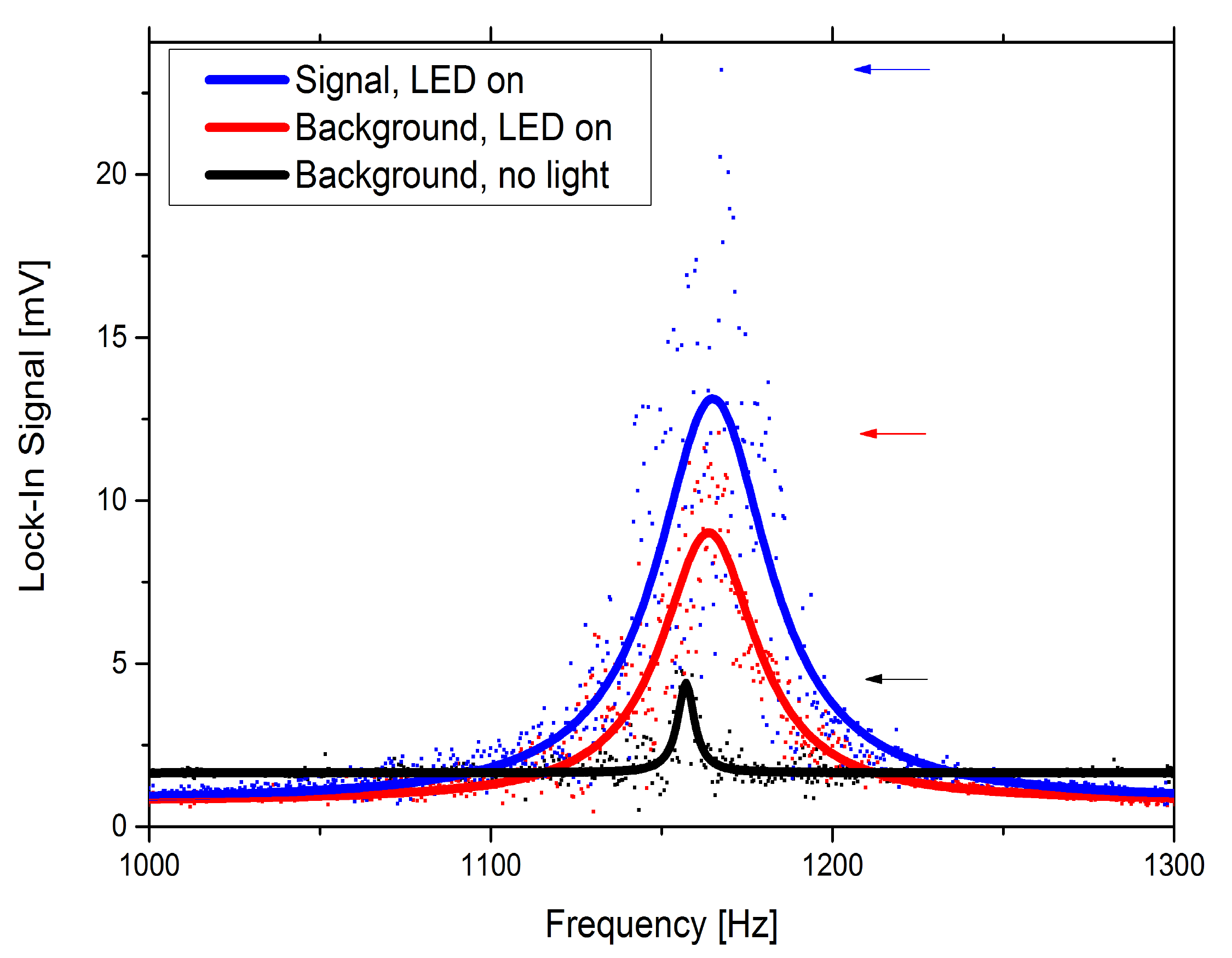}}
\caption{The lock-in response for 100 ppm concentration of NO$_2$ gas in the PA cell with the pump light on (blue), without NO$_2$ but still with pump light (red), and finally without gas or pump light (black). The points are experimental data and the solid lines are Lorentzian fits to the data. The arrows indicate the peak value of each of the data sets. The cell was evacuated for the two measurements without NO$_2$.}
\label{fig5}
\end{figure}

A demonstration of PA detection of a 100 ppm NO$_2$ concentration in synthetic air is presented in Fig.~\ref{fig5}. It is well known that NO$_2$ is a toxic gas and a regulated air pollutant. It is therefore very important to perform reliable measurements of this gas, for example in connection with hot exhaustion from engines and factories. The NO$_2$ molecule has a strong and broad absorption spectrum covering the 250-650 nm spectral region \cite{Saarela2011,lassen2014}. Above 415 nm approximately 90$\%$ of the absorbed light is converted to heat/pressure through the PA effect. We use an LED as the pump source. The LED has a wavelength of 455 nm and a 13 nm bandwidth. The data is processed using a digital lock-in amplifier with 50 ms integration time. The LED modulation is controlled by the lock-in amplifier. The peak-to-peak modulation is 130 mW and approximately 50 mW is coupled into the PA cell. The experimental data shown in Fig. \ref{fig5} demonstrates a clear difference between the PA signals generated with and without NO$_2$ inside the cell. Most likely, the signal seen when there is no NO$_2$ present in the cell is from the fiber tip oscillating without any the presence of any photoacoustic signal. Due to the fact that LED light is hard to focus, when the LED is on (but without NO2), some of the light will hit the fiber tip, which then would heat up the fiber, causing it to oscillate. Since the LED was chopped at the eigen frequency of the fiber, the fiber will oscillate at this frequency, even with no gas in the cell. Since the background signal can be stable over a long period, a practical approach for background elimination is simply to use the nonzero background as a baseline reference \cite{Szabo2013}. The minimum detectable NO$_2$ concentration can therefore simply be estimated by taking the ratio between the maximum NO$_2$ signal and the background signal level. However, judging from the spread in data points compared to the peak value, the SNR is still only around 2, leaving the minimum detectable NO$_2$ concentration at approximately 50 ppm. This relatively low SNR is due to the fact that no focusing of the LED was used and therefore the whole cell and fiber was illuminated by the LED. The SNR can thus be enhanced by using better focusing, higher optical power, and by using a proper designed resonant PA cell where the acoustic resonance of the PA cell is matched to the fiber resonance.

\section{Conclusion}

An all-optical detection method was demonstrated for the detection of acoustic pressure waves. The absence of a traditional microphone makes the demonstrated system less susceptible to the effects that a hazardous environment might have on the sensor. We have further demonstrated that the sensor is useful for measurements in environments at high temperatures up to at least $200^{\circ}$C where standard microphones might not operate. Also since the optical design and alignment of the fiber sensor is very easy and cheap it makes the sensor very attractive for commercial applications. The proof-of-concept of the all-optical detection method was demonstrated by detecting sound waves generated by the PA effect of NO$_2$ excited by illuminating the molecules with a 455 nm LED. A minimum detectable NO$_2$ concentration of approximately 50 ppm was demonstrated. Currently, experiments are being conducted in order enhance the SNR by better focusing of the pump source (laser/LED) into the cell. Also, by using a proper designed resonant PA cell a much higher SNR is expected. For this many different approached has been demonstrated aiming at various aspects of SNR \cite{Koskinen2008,lassen2014,Saarela2011}. The presented proof-of-concept experiment demonstrates that the fiber cantilever has potential as a compact and robust sound transducer for PA sensing of various trace gasses.

For future work we believe that the proposed system has the potential for miniaturization and integration for an on-chip based PA sensor. For a high sensitivity PA cell, the light source modulation has to match the acoustic resonance frequency. As the size of the PA cell is decreased the acoustic resonance frequency therefore increases, but to detect the PA signal it must always be kept much slower than the collisional relaxation process for the active medium. For the system demonstrated here this constraint is not a problem since the acoustic resonance frequency is associated with the fiber length and not the size of the PA cell.

\section{Funding Information}
We acknowledge the financial support from the Danish Agency for Science Technology and Innovation.

\section{References}


\end{document}